# Singlet axial-vector coupling constant of the nucleon in QCD without instantons

**Janardan Prasad Singh**

Physics Department, Faculty of Science,
The M. S. University of Baroda, Vadodara-390002, India

**ABSTRACT**

We have analyzed axial-vector current-current correlation functions between one-nucleon states to calculate the singlet axial-vector coupling constant of the nucleon. The octet-octet and the octet-singlet current correlators, investigated in this work, do not require any use of instanton effects. The QCD and hadronic parameters used for the evaluation of correlators have been varied by (10 - 20)%. The value of the singlet axial-vector coupling constant of the nucleon obtained from this analysis is consistent with its current determination from experiments and QCD theory.

**1. Introduction**

Knowledge of axial-vector coupling constants of the nucleon has a crucial role in understanding its longitudinal spin structure [1-4]. In a generalization of Goldberger-Treiman relation, poorly determined pseudo scalar couplings of the nucleon $g_{\eta NN}$ and $g_{\eta' NN}$ are related to the singlet coupling constant $g_A^0$ and the eighth component of SU(3) octet coupling constant $g_A^8$ [5-7]. Among the three flavor-diagonal coupling constants $g_A^a$ $(a = 3,8,0)$, the isovector coupling constant $g_A^3$ is the best understood and is measured from nuclear $\beta$-decay. The eighth component $g_A^8$ is determined from the analysis of hyperon $\beta$-decay in SU(3)$_f$ symmetry limit. Indeed, in terms of SU(3)$_f$ parameters F and D, these two axial coupling constants are expressed as

$$g_A^3 = F + D, \quad g_A^8 = 3F - D$$

and determined to be as [8,9]

$$g_A^3 = 1.270 \pm 0.003, \quad g_A^8 = 0.58 \pm 0.03$$

However, SU(3)$_f$ symmetry may be badly broken and an error in $g_A^8$ from 10% [10] to 20% [11] has been suggested. There is no direct way to measure $g_A^0$. Theoretically, its calculation is challenging on account of its association with chiral anomaly. The first moment of spin-dependent structure function g$_1$ of the nucleon can be related to the scale-invariant axial-vector coupling constants $g_A^a$ $(a=3,8,0)$ of the target nucleon. The experimental value of $g_A^0$ is obtained from measurement of g$_1$ and combining its first moment integral with the measured values of $g_A^3$ and $g_A^8$ and theoretical calculation of the perturbative QCD Wilson coefficients. Using SU(3)$_f$ symmetric value for $g_A^8$ and with no leading twist subtraction in the dispersion relation for polarized photon-nucleon scattering, COMPASS found[12]

$$g_A^0 = 0.33 \pm 0.03(stat.) \pm 0.05(syst.)$$

Several approaches have been used to calculate axial-vector coupling constants of the nucleon. Instantons, through axial anomaly relation, is believed to have an important role in the singlet axial-vector coupling constant of the nucleon [13]. Using numerical simulations of instanton liquid, Schaffer and Zetocha [14] have calculated axial-vector coupling constants of the nucleon. Though, they get a good result for $g_A^3$, for the singlet case they get $g_A^0 = 0.77$. Using lattice QCD, Yang et al. have estimated the part of the proton spin carried by light quarks from anomalous Ward identities as $\Sigma = 0.30(6)$ [15]. It hints to suggest that the culprit of the 'proton spin crisis' is the U(1)$_A$ anomaly. Chiral constituent quark model also gives a good result for $g_A^3$ and $g_A^8$, but for the singlet case, it gives $g_A^0 \simeq 0.52$ [16]. In a hybrid approach, where one takes into account one gluon exchange as well as effect of meson cloud, it has been possible to get a reasonably good result such as $g_A^0 = 0.42$ [17]. Similar result for quark spin contribution to the spin of the nucleon has been obtained using a spin-flavor based parametrization of QCD [18]. Three different approaches have been followed in QCD sum rule to calculate axial coupling constant of the nucleon. Ioffe and Oganesian [19] have used the standard QCD sum rule in external fields. Two-point correlation function of nucleon interpolating fields has been evaluated in the presence of a weak axial vector field. The limits on $\Sigma$, the part of proton spin carried by light quarks, and $\chi'(0)$, the derivative of the QCD susceptibility have been found from self-consistency of the sum rule. Belitsky and Teryaev [20] considered a three-point function of nucleon interpolating fields and the divergence of singlet axial-vector current. The form factor $g_A^0(q^2)$ is related to vacuum condensates of quark-gluon composite operators through a double dispersion relation. In this approach, the extrapolation to $g_A^0(0)$ involves large uncertainties. In the third approach by Nishikawa et al. [21,22], a two-point correlation function of axial-vector currents in one-nucleon state is evaluated. Here, the axial-vector coupling constants of the nucleon are expressed in terms of $\pi$-N and K-N sigma terms and moments of parton distributions. The perturbative contribution is subtracted from the beginning and the continuum contribution can be reduced to a small value. The application of this method using singlet-singlet axial-vector current correlator for $g_A^0$ requires taking into account the chiral anomaly [21]. This gives appreciably high value of $g_A^0 \approx 0.8$. The result was improved by the inclusion of instantons in the QCD evaluation of correlation function [22]. However, the result was extremely sensitive to critical instanton size and was not stabilized. Our own experience of working with singlet-singlet axial-vector current correlator, albeit in vacuum state [23,24], is that the sum rule does not work satisfactorily even on inclusion of instanton contribution. On the other hand, octet-octet and octet-singlet axial-vector current correlators work well. Instanton contribution is not needed in these last two sum rules. In view of this, in this work we will investigate octet-octet and octet-singlet axial-vector current correlators in one-nucleon states. The results of the two sum rules can be combined to get $|g_A^8|$ and $|g_A^0|$. The numerical evaluation of the sum rules requires use of several QCD and hadronic parameters. We have also studied consequences of variation of these parameters on sum rules.

## 2. The sum rules

Following [21,22], we consider the correlation functions of axial-vector currents in one-nucleon states:

$$\Pi_{\mu\nu}^{ab}(q;P) = i\int d^4x\, e^{iqx} \left\langle T\left[ j_{\mu 5}^a(x), j_{\nu 5}^b(0) \right] \right\rangle_N; \qquad (a,b)=(8,0) \qquad (1)$$

where

$$\langle ....\rangle_N = \frac{1}{2}\sum_S \left[ \langle PS|....|PS\rangle - \langle ....\rangle_0 \langle PS|PS\rangle \right] \qquad (2)$$

$$j_{\mu 5}^8 = \frac{1}{\sqrt{6}}\left(\bar{u}\gamma_\mu\gamma_5 u + \bar{d}\gamma_\mu\gamma_5 d - 2\bar{s}\gamma_\mu\gamma_5 s\right) \qquad (3a)$$

$$j_{\mu 5}^0 = \frac{1}{\sqrt{3}}\left(\bar{u}\gamma_\mu\gamma_5 u + \bar{d}\gamma_\mu\gamma_5 d + \bar{s}\gamma_\mu\gamma_5 s\right) \qquad (3b)$$

Actually, $\Pi_{\mu\nu}^{ab}$ has two kinds of contributions: the connected and the disconnected terms. Unlike the case of singlet-singlet correlator, the disconnected terms do not contribute to octet-octet as well as to octet-singlet correlators. Hence, the instanton contribution is not needed in our calculation. Eq.(1) can be written using Lehmann representation as

$$\Pi_{\mu\nu}^{ab}(\omega,\vec{q};P) = \int_{-\infty}^{\infty} d\omega' \frac{\rho_{\mu\nu}^{ab}(\omega',\vec{q};P)}{\omega-\omega'}, \qquad (4)$$

where $\rho_{\mu\nu}^{ab}$ is the spectral function. We take Borel transform of even part in $\omega$ of both sides of Eq. (4)

$$\hat{B}\left[\Pi_{\mu\nu}^{ab}(\omega,\vec{q};P)_{even}\right] = -\int_{-\infty}^{\infty} d\omega'\,\omega'\, e^{-\omega'^2/s}\rho_{\mu\nu}^{ab}(\omega',\vec{q};P) \qquad (5)$$

where $\hat{B} = \lim_{-\omega^2\to\infty, n\to\infty, -\omega^2/n=s} \frac{(-\omega^2)^{n+1}}{n!}\left[-\frac{d}{d(-\omega^2)}\right]^n$. The nucleon matrix element of axial-vector current is given as

$$\langle P,S|J_{\mu 5}^8|P',S'\rangle = \frac{1}{\sqrt{6}}\bar{u}(P,S)\left[g_A^8(q^2)\gamma_\mu\gamma_5 + h_A^8(q^2)q_\mu\gamma_5\right]u(P',S') \qquad (6a)$$

$$\langle P,S|J_{\mu 5}^0|P',S'\rangle = \frac{1}{\sqrt{3}}\bar{u}(P,S)\left[g_A^0(q^2)\gamma_\mu\gamma_5 + h_A^0(q^2)q_\mu\gamma_5\right]u(P',S') \qquad (6b)$$

Calling $\Pi^{ab}(\omega,\vec{q}) = \Pi^{ab\mu}_{\mu}(\omega,\vec{q};M,\vec{0})$ and realizing that $h^a_A(q^2)$ has no singularity at $q^2 = 0$, one gets

$$\frac{\partial}{\partial \vec{q}^2}\hat{B}\left[\Pi^{88}(\omega,\vec{q})\right]_{\vec{q}^2=0} = -\frac{1}{2}\frac{1}{M}|g^8_A|^2 \tag{7a}$$

$$\frac{\partial}{\partial \vec{q}^2}\hat{B}\left[\Pi^{80}(\omega,\vec{q})\right]_{\vec{q}^2=0} = -\frac{1}{\sqrt{2}}\frac{1}{M}g^8_A g^0_A \tag{7b}$$

We have calculated correlation functions $\Pi^{ab}_{\mu\nu}$ using operator product expansion (OPE) by accounting for operators up to dimension 6. Our results for $\Pi^{88}(q^2)$ has some differences from those obtained in Ref.[21].

$$\begin{aligned}\Pi^{88}(q^2) &= \frac{1}{6}\Bigg[\frac{10}{q^2}\left(m_u\langle\bar{u}u\rangle_N + m_d\langle\bar{d}d\rangle_N + 4m_s\langle\bar{s}s\rangle_N\right) - \frac{3}{2}\frac{1}{q^2}\left\langle\frac{\alpha_s}{\pi}G^2\right\rangle_N \\ &\quad -8\frac{q^\mu q^\nu}{q^4}i\langle\bar{u}S\gamma_\mu D_\nu u + \bar{d}S\gamma_\mu D_\nu d + 4\bar{s}S\gamma_\mu D_\nu s\rangle_N \\ &\quad -\frac{3}{2}\frac{\pi\alpha_s}{q^4}\Bigg[\left\langle(\bar{u}\gamma_\mu\lambda^a u + \bar{d}\gamma_\mu\lambda^a d + 4\bar{s}\gamma_\mu\lambda^a s)\sum_i \bar{q}_i\gamma^\mu\lambda^a q_i\right\rangle_N \\ &\quad +4\langle(\bar{u}\gamma_\mu\lambda^a u + \bar{d}\gamma_\mu\lambda^a d - 2\bar{s}\gamma_\mu\lambda^a s)^2\rangle_N\Bigg] \\ &\quad -\frac{2}{3}\frac{\pi\alpha_s}{q^4}\frac{q^\mu q^\nu}{q^2}\Bigg[\left\langle S((\bar{u}\gamma_\mu\lambda^a u + \bar{d}\gamma_\mu\lambda^a d + 4\bar{s}\gamma_\mu\lambda^a s)\sum_i \bar{q}_i\gamma_\nu\lambda^a q_i)\right\rangle_N \\ &\quad -12S\langle(\bar{u}\gamma_\mu\lambda^a u + \bar{d}\gamma_\mu\lambda^a d - 2\bar{s}\gamma_\mu\lambda^a s)(\mu\to\nu)\rangle_N\Bigg] \\ &\quad +32i\frac{q^\beta q^\rho q^\lambda q^\sigma}{q^8}\langle\bar{u}S\gamma_\beta D_\rho D_\lambda D_\sigma u + \bar{d}S\gamma_\beta D_\rho D_\lambda D_\sigma d + 4\bar{s}S\gamma_\beta D_\rho D_\lambda D_\sigma s\rangle_N\Bigg]\end{aligned} \tag{8a}$$

$$\begin{aligned}\Pi^{80}(q^2) &= \frac{1}{3\sqrt{2}}\Bigg[\frac{10}{q^2}\left(m_u\langle\bar{u}u\rangle_N + m_d\langle\bar{d}d\rangle_N - 2m_s\langle\bar{s}s\rangle_N\right) \\ &\quad -8\frac{q^\mu q^\nu}{q^4}i\langle\bar{u}S\gamma_\mu D_\nu u + \bar{d}S\gamma_\mu D_\nu d - 2\bar{s}S\gamma_\mu D_\nu s\rangle_N \\ &\quad -\frac{3}{2}\frac{\pi\alpha_s}{q^4}\Bigg[\left\langle(\bar{u}\gamma_\mu\lambda^a u + \bar{d}\gamma_\mu\lambda^a d - 2\bar{s}\gamma_\mu\lambda^a s)\sum_i \bar{q}_i\gamma^\mu\lambda^a q_i\right\rangle_N \\ &\quad +4\left\langle(\bar{u}\gamma_\mu\lambda^a u + \bar{d}\gamma_\mu\lambda^a d - 2\bar{s}\gamma_\mu\lambda^a s)\sum_i \bar{q}_i\gamma^\mu\lambda^a q_i\right\rangle_N\Bigg] \\ &\quad -\frac{2}{3}\frac{\pi\alpha_s}{q^4}\frac{q^\mu q^\nu}{q^2}\left[\left\langle S((\bar{u}\gamma_\mu\lambda^a u + \bar{d}\gamma_\mu\lambda^a d - 2\bar{s}\gamma_\mu\lambda^a s)\sum_i \bar{q}_i\gamma_\nu\lambda^a q_i)\right\rangle_N\right.\end{aligned} \tag{8b}$$

$$-12\left\langle S\left[(\bar{u}\gamma_\mu\lambda^a u+\bar{d}\gamma_\mu\lambda^a d-2\bar{s}\gamma_\mu\lambda^a s)\sum_i \bar{q}_i\gamma_\nu\lambda^a q_i\right]\right\rangle_N$$

$$+32i\frac{q^\beta q^\rho q^\lambda q^\sigma}{q^8}\left\langle \bar{u}S\gamma_\beta D_\rho D_\lambda D_\sigma u+\bar{d}S\gamma_\beta D_\rho D_\lambda D_\sigma d-2\bar{s}S\gamma_\beta D_\rho D_\lambda D_\sigma s\right\rangle_N\Big]$$

where S stands for symmetrization in Lorentz indices and summation is over flavors u,d and s. The quark matrix elements $m_q\langle\bar{q}q\rangle_N$ (q=u,d,s) are expressed in terms of $\pi$-N and K-N sigma terms and the gluonic matrix element $\langle(\alpha_s/\pi)G^2\rangle_N$ is expressed in terms of nucleon mass and $m_q\langle\bar{q}q\rangle_N$ through the QCD trace anomaly. $y=2\langle\bar{s}s\rangle_N/\langle\bar{u}u+\bar{d}d\rangle_N$ is a measure for the strange quark content of the nucleon. The matrix elements containing covariant derivatives are related to the parton distributions as $\left\langle S\bar{q}\gamma_{\mu_1}D_{\mu_2}...D_{\mu_n}q(\mu^2)\right\rangle_N=(-i)^{n-1}A_n^q(\mu^2)S[P_{\mu_1}P_{\mu_2}...P_{\mu_n}]$, where $A_n^q(\mu^2)$ is the n$^{th}$ moment of parton distribution of q-type parton. Also matrix elements of four-quark operators have been factorized assuming dominance of one-nucleon state as the intermediate state: $\langle O_1 O_2\rangle_N \approx \langle O_1\rangle_N\langle O_2\rangle_0+\langle O_1\rangle_0\langle O_2\rangle_N$. From Eqs.(7a,b) and (8a,b), we obtain isospin averaged expression for $g_A^8$ and $g_A^0$ as

$$\left|g_A^{(8)}\right|^2=-\frac{M}{3}\left[\frac{\Sigma_{\pi N}}{3s}\left(26-\frac{116m_s}{m_u+m_d}\right)+\frac{\Sigma_{KN}}{3s}\frac{232m_s}{m_s+m_u}+\frac{M}{s}\left\{\frac{4}{3}-7\left(A_2^u+A_2^d+4A_2^s\right)\right\}\right.$$
$$\left.-\frac{4\pi\alpha_s}{s^2}\frac{40}{3}\langle\bar{q}q\rangle_0\frac{\Sigma_{\pi N}}{m_u+m_d}+\frac{30M^3}{s^2}\left(A_4^u+A_4^d\right)-\frac{4\pi\alpha_s}{s^2}\frac{80}{3}\langle\bar{s}s\rangle_0\frac{y\Sigma_{\pi N}}{m_u+m_d}\right] \quad (9a)$$

$$g_A^{(0)}g_A^{(8)}=-\frac{M}{3}\left[\frac{10}{s}\left(\frac{-4m_s}{m_u+m_d}\Sigma_{KN}+\left(1+\frac{m_s}{m_u}\right)\Sigma_{\pi N}\right)-7\frac{M}{s}\left(A_2^u+A_2^d-2A_2^s\right)\right.$$
$$\left.-\frac{4\pi\alpha_s}{s^2}\frac{40}{3}\langle\bar{q}q\rangle_0\frac{\Sigma_{\pi N}}{m_u+m_d}+\frac{30M^3}{s^2}\left(A_4^u+A_4^d\right)+\frac{4\pi\alpha_s}{s^2}\frac{40}{3}\langle\bar{s}s\rangle_0\frac{y\Sigma_{\pi N}}{m_u+m_d}\right] \quad (9b)$$

In above equations we have not taken into account continuum contribution about which we will comment later. It may be pointed out that the last terms in Eqs. (9a,b) arising from $\langle\bar{s}\gamma_\mu\lambda^a s\bar{s}\gamma^\mu\lambda^a s\rangle_N$ has not been considered earlier [21,22]. Our expression on the rhs of Eq.(9a) differs from the corresponding expression in Ref.[21] in other significant ways: the second and third terms differ in sign whereas fourth and fifth terms have somewhat different numerical coefficients.

## 3. Results and Discussion

In the current literature, there are significant variations in the values of the QCD and hadronic parameters appearing in Eqs. (9a) and (9b). The most important parameters are the second moments of the parton distributions. For calculating $A_n^q(\mu^2)$, we have used MSTW 2008 parametrization of parton distributions at $\mu = 1 GeV$ [25]. This gives $A_2^u + A_2^d = 0.9724, A_2^s = 0.0479$, $A_4^u + A_4^d = 0.1206$ and $A_4^s = 0.0011$. These authors have also used NNLO parametrization of strong coupling constant at $\mu = 1 GeV$ as $\alpha_s(\mu) = 0.45077$ whereas it is common to use $\alpha_s(\mu) = 0.5$ in QCD sum rule calculations [23,24]. For 2+1 flavors from lattice QCD world data, Alvarez-Ruso et al.[26] have determined $\Sigma_{\pi N} = \frac{1}{2}(m_u + m_d)\left(\langle \overline{u}u \rangle_N + \langle \overline{d}d \rangle_N\right) = 52(3)(8) MeV$. For K-N sigma term $\Sigma_{KN} = \frac{1}{2}(m_q + m_s)\left(\langle \overline{u}u \rangle_N + \langle \overline{s}s \rangle_N\right)$, Nowak, Rho and Zahed [13] estimate $\Sigma_{KN} \approx (2.3, 2.8, 3.4)m_\pi$ for y=0, 0.1 and 0.2 with $m_q = \frac{1}{2}(m_u + m_d) = 5 MeV, m_s = 135 MeV$ and $\Sigma_{\pi N} = 45 MeV$. Actually, for the strange quark content of the nucleon, the lattice result for y is considered more accurate than sigma-term and is given as y=0.135(22)(33)(22)(9) [27]. For current quark masses at $\mu = 1 GeV$, Ioffe et al. [28] have obtained $m_s \approx 147 MeV$ assuming $m_s(2GeV) = 120 MeV$, and $m_u + m_d = 10.0 \pm 2.5 MeV$ whereas $m_u + m_d = 12.8 \pm 2.5 MeV$ in Ref.[29]. For light quark vacuum condensate $\langle \overline{q}q \rangle_0 = \langle \overline{u}u \rangle_0 = \langle \overline{d}d \rangle_0$, value for $(2\pi)^2 \langle \overline{q}q \rangle_0$ between 0.45 GeV³ to 0.65 GeV³ has been commonly used [23,24,28].

In view of this prevailing uncertainty in numerical values of these parameters, it is desirable to take these uncertainties into account while determining the axial coupling constants of the nucleon from Eqs. (9a,b). We will vary each of these parameters by (10-20)% of their certain central values covering roughly their ranges as given above with an aim to obtain values of axial coupling constants of the nucleon as determined by experiments and phenomenological analyses. In addition to this, we also chose $\langle \bar{s}s \rangle_0 = 0.8 \langle \bar{q}q \rangle_0$ [28], and $A_2^s = 0.048$ from MSTW 2008 [25]. We observed that the sum rules were giving unphysical results for $g_A^8$ and $g_A^0$ for Borel mass squared $s \leq 1.3 GeV^2$ and reasonable results are obtained for $s \geq 1.6 GeV^2$ for various combinations of QCD and hadronic parameters. We seek stable results for $|g_A^8|$ and $|g_A^0|$ in the range $1.7 GeV^2 \leq s \leq 2.5 GeV^2$. For this, we first varied $m_s, m_q, \Sigma_{\pi N}, \Sigma_{KN}, \langle \bar{q}q \rangle_0, \alpha_s, A_2^u + A_2^d, A_4^u + A_4^d$ and y by ~(10-20)% around a central value of each of these parameters as given in the second row of Table 1. First we found those sets of parameters for which $0.52 < |g_A^8| < 0.64$ is obtained from Eq.(9a) for $1.7 GeV^2 \leq s \leq 2.5 GeV^2$. These parameter sets were further constrained by requiring that $|g_A^0|$ at the middle of the Borel mass parameter range, i.e. at $s = 2.1 GeV^2$, lies in the range 0.29-0.42. Among these sets of parameters, we looked for those which were giving most stable results for $|g_A^8|$ and $|g_A^0|$ against variation of Borel mass squared parameter s. Actually, $\Sigma_{KN}$ and y are not independent parameters and are related as $\Sigma_{KN}(y) = \frac{1}{4}(1+y) \times \left(1 + \frac{m_s}{m_q}\right) \Sigma_{\pi N}$. We found that for a given set of $m_s, m_q$ and $\Sigma_{\pi N}$, if $\Sigma_{KN}$ and y are chosen according to this relation then the sum rules do not work well. Hence we have varied $\Sigma_{KN}$ and y independently while y has been used only in the last terms of Eqs.(9a) and (9b). However, we have tried to keep $\Sigma_{KN}$ and $\Sigma_{KN}(y)$ as close as possible and maintained $\Sigma_{KN}$ to be in the range of (82-85)% of $\Sigma_{KN}(y)$. In Table 1, only those results are displayed for which $\Sigma_{KN}$ and $\Sigma_{KN}(y)$ are closest possible for a given set of $m_s, m_q$ and $\Sigma_{\pi N}$. We may consider this use of independent values for $\Sigma_{KN}$ and y or $\Sigma_{KN}(y)$ as a way to compensate the possible violation of factorization hypothesis used in the last terms of Eqs. (9a) and (9b). We get a wide range of combination of parameters $m_s, m_q, \Sigma_{\pi N}, \Sigma_{KN}, \langle \bar{q}q \rangle_0, \alpha_s, A_2^u + A_2^d, A_4^u + A_4^d$ and y being used in the current literature for which the sum rules give values of $|g_A^8|$ and $|g_A^0|$ which lie in the typical range that is obtained from experimental and phenomenological analyses. We believe this as a sign of robustness of our sum rules. In Table 1 we have listed some of those results for which $|g_A^0|$, as a function of s, has minimum slopes in our designated interval s= (1.7-2.5) GeV². Plots of some of these results are displayed in Figs. (1-6). We observe from the Table 1 that $A_2^u + A_2^d$ was stuck to the lower end of the range of its variation and $\Sigma_{KN}$ was confined to (300-325) MeV. The best results were obtained for y being in the range (0.16 - 0.18).

As in any QCD sum rule calculation, our results have errors due to omission of contributions of higher dimensional operators and continuum contributions. From MSTW 2008 parametrization [25], we estimate $A_3^u + A_3^d \simeq 0.3, A_6^u + A_6^d \simeq 0.03$ and $(A_3^u + A_3^d)\langle \bar{q}q \rangle_0 \simeq 3.8 \times 10^{-3} GeV^3$. The ratio of contributions of six-dimensional operators to that of four-dimensional operators is ~1/2 at s=2.5 GeV², but the ratio of contribution of eight-dimensional operators to that of four-dimensional operator is likely to be few percent, though their contribution to $|g_A^{8,0}|$ will get doubled on account of sign difference in the contributions of four-dimensional and six-dimensional operators. The continuum contribution comes from $\eta - N$ and $\eta' - N$ states. A rough estimate shows that their contribution will be less than 1%. Thus we allow the error due to exclusion of contributions of higher dimensional operators and continuum contributions to be roughly 10%. Based on results given in Table 1 and the error estimates, we conclude

$$|g_A^8| = 0.59 \pm 0.05 \pm 0.06 \tag{10a}$$

$$|g_A^0| = 0.39 \pm 0.05 \pm 0.04 \tag{10b}$$

where the first error is due to finite slope within the designated range of Borel mass parameter and the second one is due to omission of contributions of higher dimensional operators and continuum contribution.

By choosing the correlator of singlet and octet axial-vector currents, we ensured that the disconnected diagrams do not contribute directly for determination of $g_A^0$. However, the non-valence components in the nucleon, such as strange quark-antiquarks and gluons, have an important role: they are directly responsible for the splitting of $g_A^8$ and $g_A^0$. In QCD parton model, the axial coupling constants of a nucleon are related to polarized quark densities. Our results for $g_A^8$ and $g_A^0$ implies that polarized s-quark density is negative: $\Delta s \sim -0.08$. A numerical analysis of Eqs. (8a-9b) shows that $\Delta s$ gets contribution from $\langle \bar{s}s \rangle_N, A_2^s, \langle \frac{\alpha_s}{\pi} G^2 \rangle_N$ and $\langle \bar{s}s \rangle_0 y$. While the first two quantities make $\Delta s$ negative with $A_2^s$ giving dominant contribution, the last two quantities contribute positively with the gluon contribution being dominant one. The use of four s-quark operators and its subsequent evaluation in the form of $\langle \bar{s}s \rangle_0 y$ term by factorization hypothesis gives a semblance of "disconnected" diagram contributing to sum rules. However, this contribution is the smallest one. We can also look at the problem of negative $\Delta s$ using the generalized GT relation [7]. Realizing that $\bar{s}\gamma_\mu\gamma_5 s = (j_{\mu 5}^0 - \sqrt{2} j_{\mu 5}^8)/\sqrt{3}$, we can define its decay constants as $\langle 0|\bar{s}\gamma_\mu\gamma_5 s|\eta^{()}(p)\rangle = ip_\mu f_{\eta^{()}}^s$ and estimate $f_\eta^s \simeq -124.2 MeV$ and $f_{\eta'}^s \simeq 115.0 MeV$ from $f_{\eta^{()}}^{0,8}$ [23,24]. Also from $g_{\eta NN}$ =(3 - 5) [30] and $g_{\eta' NN}$ =(1 - 2) [7], and on using U(1)$_A$ GT relation for s-quark only gives $-\Delta s \simeq (0.09 - 0.16)$.

## 4. Conclusion

By considering the correlation function of octet-octet and octet-singlet axial-vector currents between one-nucleon states, we have obtained sum rules for $|g_A^8|^2$ and $g_A^0 g_A^8$ without use of instantons. For numerical evaluation of $|g_A^8|$ and $|g_A^0|$, we use sets of QCD and hadronic parameters which appear in our sum rules such that they lie in a range which has been obtained from phenomenological and theoretical analyses in recent years. We found that there exists a large number of such parameter sets which can yield $|g_A^8|$ and $|g_A^0|$ that lie within a range which is consistent with the current determination of their values from experimental and phenomenological analyses. Basically, we chose QCD and hadronic parameter sets which yield, through our sum rules, values of $|g_A^8|$, in a chosen interval of Borel mass parameter, in a range which is phenomenologically acceptable. We further restricted these sets of parameters so that the values of $|g_A^0|$ at the middle of the Borel mass squared parameter interval lie in a range which is currently acceptable from experimental data combined with theoretical QCD analysis.

We accept $|g_A^0|$ obtained in the entire designated interval of Borel mass parameter as our final result of sum rules. We also note an interesting point that the sign of spin-dependent parton density $\Delta s$ is decided by spin-independent quantities such as second moment of spin-independent parton distribution function of s-type and s-quark content of the nucleon. In conclusion, the present method of QCD sum rule, where correlation function of two axial-vector currents between one-nucleon states is studied, is capable of producing a result for singlet axial-vector coupling constant of the nucleon which is consistent with its current determination from experiments and QCD theoretical analysis.

**ACKNOWLEDGEMENT** : The author thanks Department of Science & Technology, Government of India, for financial support.

Table 1: Our results for $|g_A^8|$ and $|g_A^0|$ for s= 1.7 GeV$^2$ and 2.5 GeV$^2$ with low slopes in s. $\Delta g_A^0 = |g_A^0(2.5)| - |g_A^0(1.7)|$ . $m_q, m_s, \Sigma_{\pi N}$ and $\Sigma_{KN}$ are in MeV and $-\langle \bar{q}q \rangle (2\pi)^2$ is in GeV$^3$. In the second row, the range of variation of the parameters appearing in the first row is shown.

| y | $m_s$ | $\Sigma_{\pi N}$ | $\Sigma_{KN}$ | $-\langle \bar{q}q \rangle (2\pi)^2$ | $\alpha_s$ | $A_2^u + A_2^d$ | $A_4^u + A_4^d$ | $m_q$ | $|g_A^8(1.7)|$ | $|g_A^8(2.5)|$ | $|g_A^0(1.7)|$ | $|g_A^0(2.5)|$ | $\Delta g_A^0$ |
|---|---|---|---|---|---|---|---|---|---|---|---|---|---|
| 0.13-0.20 | 135-155 | 45-60 | 250-400 | 0.45-0.60 | 0.45-0.50 | 0.95-1.00 | 0.11-0.13 | 5-7 | | | | | |
| 0.16 | 155 | 57.5 | 325 | 0.45 | 0.475 | 0.95 | 0.12 | 7 | 0.579 | 0.639 | 0.344 | 0.441 | 0.098 |
| 0.16 | 155 | 57.5 | 325 | 0.475 | 0.45 | 0.95 | 0.12 | 7 | 0.579 | 0.639 | 0.344 | 0.441 | 0.098 |
| 0.17 | 135 | 52.5 | 300 | 0.45 | 0.50 | 0.95 | 0.11 | 6 | 0.558 | 0.634 | 0.341 | 0.440 | 0.099 |
| 0.17 | 135 | 52.5 | 300 | 0.475 | 0.475 | 0.95 | 0.11 | 6 | 0.557 | 0.633 | 0.340 | 0.440 | 0.100 |
| 0.17 | 135 | 52.5 | 300 | 0.50 | 0.45 | 0.95 | 0.11 | 6 | 0.558 | 0.634 | 0.341 | 0.440 | 0.099 |
| 0.17 | 150 | 55.0 | 300 | 0.55 | 0.45 | 0.95 | 0.11 | 7 | 0.560 | 0.633 | 0.341 | 0.439 | 0.098 |
| 0.17 | 150 | 55.0 | 300 | 0.50 | 0.50 | 0.95 | 0.11 | 7 | 0.555 | 0.631 | 0.337 | 0.438 | 0.101 |
| 0.17 | 155 | 57.5 | 325 | 0.45 | 0.45 | 0.95 | 0.13 | 7 | 0.573 | 0.637 | 0.338 | 0.439 | 0.101 |
| 0.18 | 150 | 47.5 | 300 | 0.50 | 0.475 | 0.95 | 0.12 | 6 | 0.562 | 0.637 | 0.340 | 0.441 | 0.100 |
| 0.18 | 150 | 47.5 | 300 | 0.525 | 0.45 | 0.95 | 0.12 | 6 | 0.565 | 0.639 | 0.342 | 0.441 | 0.099 |
| 0.18 | 150 | 47.5 | 300 | 0.475 | 0.50 | 0.95 | 0.12 | 6 | 0.562 | 0.637 | 0.340 | 0.441 | 0.100 |
| 0.18 | 150 | 47.5 | 300 | 0.575 | 0.45 | 0.95 | 0.11 | 6 | 0.542 | 0.629 | 0.342 | 0.442 | 0.100 |
| 0.20 | 135 | 52.5 | 300 | 0.45 | 0.45 | 0.95 | 0.13 | 6 | 0.544 | 0.628 | 0.341 | 0.441 | 0.099 |
| 0.20 | 150 | 55.0 | 300 | 0.45 | 0.50 | 0.95 | 0.13 | 7 | 0.540 | 0.626 | 0.337 | 0.438 | 0.101 |
| 0.20 | 150 | 55.0 | 300 | 0.50 | 0.45 | 0.95 | 0.13 | 7 | 0.540 | 0.626 | 0.337 | 0.438 | 0.101 |

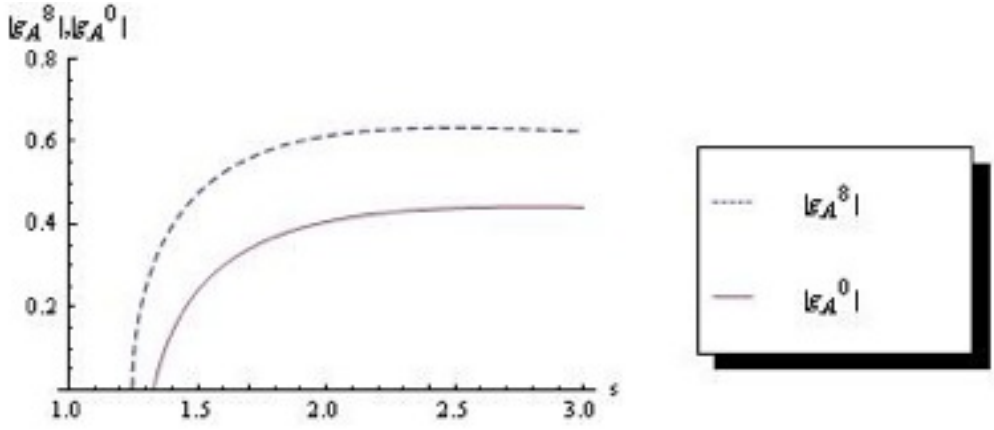

Fig.1: y=0.16, $m_s$ =155 MeV, $\Sigma_{\pi N}$ =57.5 MeV, $\Sigma_{KN}$ =325 MeV, $-\langle \bar{q}q \rangle (2\pi)^2$ =0.475 GeV³, $\alpha_s$ =0.45, $A_2^u + A_2^d$ =0.95, $A_4^u + A_4^d$ =0.12, $m_q$ =7 MeV and $\Delta g_A^0$ = 0.098.

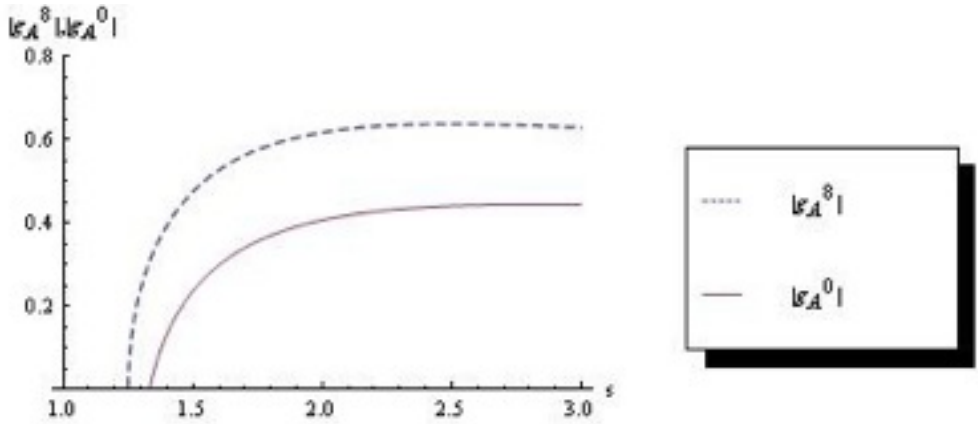

Fig.2: y=0.17, $m_s$ =135 MeV, $\Sigma_{\pi N}$ =52.5 MeV, $\Sigma_{KN}$ =300 MeV, $-\langle \bar{q}q \rangle (2\pi)^2$ =0.50 GeV³, $\alpha_s$ =0.45, $A_2^u + A_2^d$ =0.95, $A_4^u + A_4^d$ =0.11, $m_q$ =6 MeV and $\Delta g_A^0$ = 0.099.

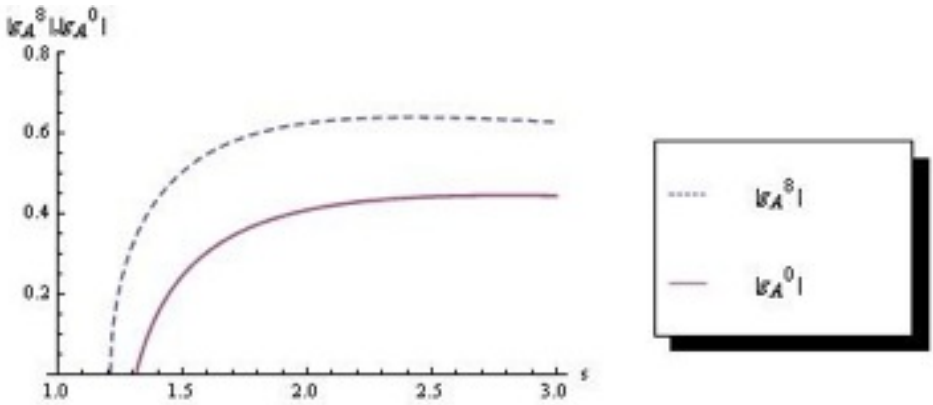

Fig. 3: y=0.17, $m_s$ =150 MeV, $\Sigma_{\pi N}$ =55.0 MeV, $\Sigma_{KN}$ =300 MeV, $-\langle \bar{q}q \rangle (2\pi)^2$ =0.55 GeV³, $\alpha_s$ =0.45, $A_2^u + A_2^d$ =0.95, $A_4^u + A_4^d$ =0.11, $m_q$ =7 MeV and $\Delta g_A^0$ = 0.098.

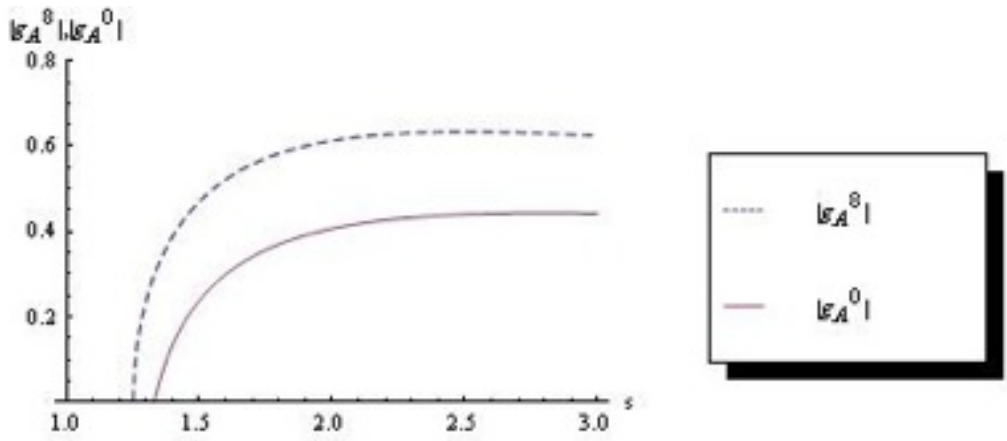

Fig. 4: y=0.18, $m_s$ =150 MeV, $\Sigma_{\pi N}$ =47.5 MeV, $\Sigma_{KN}$ =300 MeV, $-\langle \bar{q}q \rangle (2\pi)^2$ =0.50 GeV³, $\alpha_s$ =0.475, $A_2^u + A_2^d$ =0.95, $A_4^u + A_4^d$ =0.12, $m_q$ =6 MeV and $\Delta g_A^0$ = 0.100.

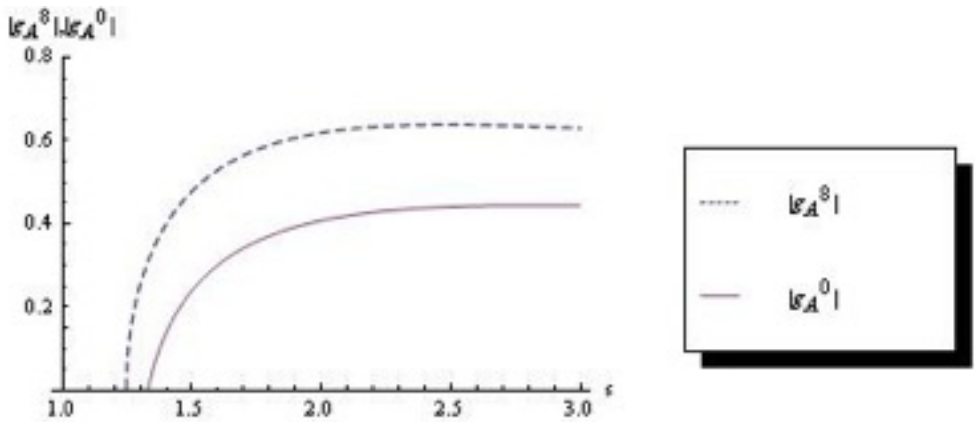

Fig.5: y=0.18, $m_s$ =150 MeV, $\Sigma_{\pi N}$=47.5 MeV, $\Sigma_{KN}$ =300 MeV, $-\langle \bar{q}q \rangle (2\pi)^2$ =0.525 GeV³, $\alpha_s$=0.45, $A_2^u + A_2^d$ =0.95, $A_4^u + A_4^d$ =0.12, $m_q$ =6 MeV and $\Delta g_A^0$= 0.099.

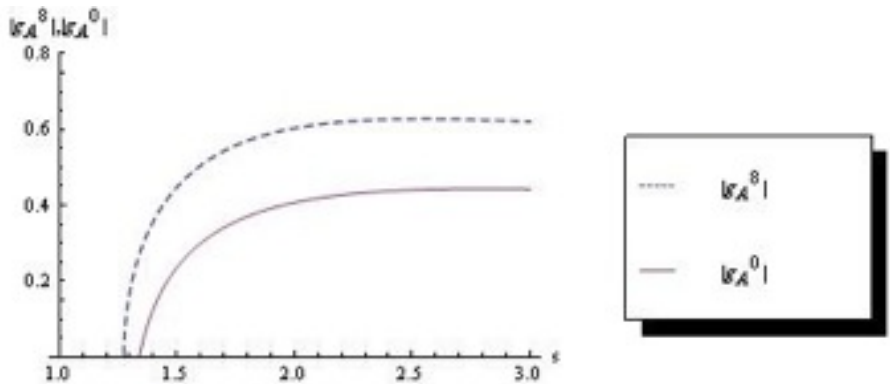

Fig.6: y=0.20, $m_s$ =135 MeV, $\Sigma_{\pi N}$ =52.5 MeV, $\Sigma_{KN}$ =300 MeV, $-\langle \bar{q}q \rangle (2\pi)^2$ =0.45 GeV³, $\alpha_s$ =0.45, $A_2^u + A_2^d$ = 0.95, $A_4^u + A_4^d$ = 0.13, $m_q$= 6 MeV and $\Delta g_A^0$ = 0.099.